\documentclass[amsmath,amssymb,superscriptaddress,showkeys, showpacs, aps, nofootinbib]{revtex4}
\usepackage{graphicx}
\usepackage{amstext}
\usepackage{mathrsfs}
\usepackage{xcolor}


\begin{document}

\vspace{0.5 in}
\markboth{R.~L.~Hall \& P. Zorin}{Sharp comparison theorems for the Klein--Gordon equation in $d$ dimensions.}

\title{Sharp comparison theorems for the Klein--Gordon equation in $d$ dimensions.}

\author{Richard L. Hall}
\email{richard.hall@concordia.ca}
\affiliation{Department of Mathematics and Statistics, Concordia University,
1455 de Maisonneuve Boulevard West, Montr\'eal,
Qu\'ebec, Canada H3G 1M8}

\author{Petr~Zorin}
\email{petrzorin@yahoo.com}
\affiliation{Department of Mathematics and Statistics, Concordia University,
1455 de Maisonneuve Boulevard West, Montr\'eal,
Qu\'ebec, Canada H3G 1M8}

\begin{abstract} 
We establish sharp (or `refined') comparison theorems for the Klein--Gordon equation. We show that the condition $V_a\le V_b$, which leads to $E_a\le E_b$, can be replaced by the weaker assumption $U_a\le U_b$ which still implies the spectral ordering $E_a\le E_b$. In the simplest case, for $d=1$, $U_i(x)=\int_0^x V_i(t)dt$, $i=a$ or $b$, and for $d>1$, $U_i(r)=\int_0^r V_i(t) t^{d-1}dt$, $i=a$ or $b$. We also consider sharp comparison theorems in the presence of a scalar potential $S$  (a `variable mass') in addition to the vector term $V$ (the time component of a $4$-vector). The theorems are illustrated by a variety  of explicit detailed examples.  
\end{abstract}



\maketitle

\section{Introduction}
There are few exact analytical solutions of the Klein--Gordon equation for bound systems. Therefore methods that give approximate solutions, particularly bounds for the solutions, can be very useful. The comparison theorem of quantum mechanics allows us to obtain upper or lower bounds for an eigenvalue with the aid of suitable comparison potentials that do have known exact solutions. Such a theorem states that if two comparison potentials are ordered, i.e. $V_a\le V_b$, then the corresponding discrete energy eigenvalues are similarly ordered $E_a\le E_b$. In the nonrelativistic case this follows directly from the min--max variational principle \cite{Reed, Thirring}. In the relativistic case, since the Hamiltonian is not bounded below, a min--max characterization is not immediatelly available. Nevertheless, by the use of other techniques, comparison theorems have been established \cite{Hall1, Hall2, monoton1, monoton2, monoton3}. 

The results for the Klein--Gordon problem to date are limited to cases where the potentials are negative and strictly ordered and the eigenvalues are nonnegative.
Counter examples to a simple general comparison theorem $V_a\le V_b$ $\Rightarrow$ $E_a\le E_b$ for the Klein--Gordon equation are immediately provided by (1) the square--well potential shape in $d=1$ dimension, $f(x)=-1$, $|x|\le a$, $f(x)=0$, $|x|>1$, where $V(x)=vf(x)$ for which the corresponding eigenvalues $E(v)$ are not monotonic \cite{Greiner} in $v$ when $E(v)<0$, and (2) the cut--off Coulomb potential $V(r)=-v/(a+r)$ in $d=3$ dimensions, which yields eigenvalues $E(a,\ v)$ that are not monotone in $a$ when $E(a,\ v)<0$ \cite{Barton, HallCO}. 

If we consider positive eigenvalues of problems with negative potentials, we are able to derive sharp (or `refined') comparison theorems which allow the graphs of the comparison potentials to cross over in a controlled fashion and still imply definite ordering of the respective ground--state eigenvalues for $d=1$, or at the bottom of an angular momentum subspace for $d>1$. The comparison theorem was first refined by Hall in the nonrelativistic case for $d=1$ and $d=3$ dimensions \cite{Hall3} and in $d>1$ dimensions \cite{ddimSch}, and then applied to Sturm--Liouville problems in \cite{HallComp, Hall4}. The principal aim of the present paper is to derive such sharp comparison theorems for the relativistic Klein--Gordon spectral problem. In the simplest case in one dimension we able to conclude $E_a\le E_b$ from the weaker potential assumption $U_a\le U_b$, where $U_i(x)=\int_0^x V_i(t)dt$, $i=a$ or $b$. If one of the wave functions is known, i.e. either $\varphi_a$ for the potential $V_a$ or $\varphi_b$ for $V_b$, then $U_a\le U_b$ $\Rightarrow$ $E_a\le E_b$, where $U_i(x)=\int_0^x V_i(t)\varphi_jdt$, $i=a$ or $b$ and $j$ is either equals t$a$ or $b$.  

The theorems we are able to derive are limited to the ground--state or (for $d>1$) to the bottom of an angular momentum subspace labelled by $l$. The equivalence theorem $3$ states that $E_{nl}^d=E_{n0}^{d+2l}$, where $ E_{nl}^d $ is the $n^{\text{th}}$ discrete eigenvalue in the angular-momentum space labelled by $l$ in $d>1$ dimensions. Thus by choosing $d$ large enough we can extend our results from $l=0$ to the $l>0$ cases. For instance, if $d=6$ and $l=0$ we can apply that comparison result to the family of equivalent comparison problems: $d=2$ and $l=2$ or $d=4$ and $l=1$.

\section{Sharp comparison theorems in $d=1$ dimension}
The Klein--Gordon equation in one dimension is given for example in Ref.~\cite{KGd1_1, Greiner}
\begin{equation}\label{KG1}
\varphi ''=\left[m^2-(E-V)^2\right]\varphi,
\end{equation}
where natural units $\hbar=c=1$ are used  and $E$ is the energy of a spinless particle of mass $m$. We shall assume that the potential function $V$ satisfies 
\begin{eqnarray*}
&1^\circ&V\ \text{is even, i.\,e.} \  V(x)=V(-x);\\
\vspace{9mm}
&2^\circ&V\ \text{is nonpositive, i.\,e.} \  V\le 0;\\
\vspace{9mm}
&3^\circ&V\  \text{is attractive, that is to say monotone nondecreasing on} \  [0, \infty);\\
\vspace{9mm}
&4^\circ&V\ \text{vanishes at infinity, thus}\ \lim_{x\to\pm\infty}V=0.
\end{eqnarray*}
We assume that $V$ is in this class of potentials, that at least one discrete energy eigenvalue $E$ exists, and that equation (\ref{KG1}) is the 
eigenequation for the corresponding eigenstates. According to $4^\circ$, 
equation (\ref{KG1}) at infinity becomes
\begin{equation}\label{KG1inf}
\varphi ''=(m^2-E^2)\varphi,
\end{equation} 
with solution in the form $\varphi=c_1e^{\sqrt{k}|x|}+c_2e^{-\sqrt{k}|x|}$, where 
$c_1$ and $c_2$ are constants of integration, and $k=m^2-E^2$. 
The radial wave function $\varphi$ has to vanish at infinity, thus $c_1=0$ 
and for the large--$|x|$ asymptotic form of $\varphi$ is $\varphi=c_2e^{-\sqrt{k}|x|}$. Since $\varphi\in L^2(\mathbb{R})$, $k>0$ which is equivalent to $-m<E<m$. Parenthetically, we note that similar reasoning yields the same spectral bouds in $d>1$ dimensions. It will be clear later that we shall need  to consider only cases where $0\le E<m$.

Suppose that $\varphi(x)$ is a solution of (\ref{KG1}). Then, by direct substitution, we find that $\varphi(-x)$ is also solution of (\ref{KG1}) with the same energy $E$. Therefore, by using linear combinations, we see that the eigenfunctions of (\ref{KG1}) can be taken to be even or odd. Since we consider only ground states, an odd wave function has to be excluded because it has a node at $x=0$. Thus for the present discussion $\varphi$ is an even function $\varphi\in L^2(\mathbb{R})$ but unnormalized so
\begin{equation*}
||\varphi||^2=\int_{-\infty}^\infty \varphi^2dx<\infty.
\end{equation*}
Also without loss of generality, we put $\varphi(0)=1$. 

First we prove the following lemma, which characterizes the behaviour of the nodeless wave function $\varphi$:

\medskip

\noindent{\bf Lemma 1:} ~~{\it The Klein--Gordon ground--state energy eigenfunction $\varphi$ is nonincreasing, i.e.
\begin{equation*}
\varphi'\le 0, \quad x\in[0,\ \infty).
\end{equation*}} 

\medskip

\noindent{\bf Proof:} If the potential $V$ is unbounded near the origin, i.\,e. $\lim\limits_{x\to 0}V=-\infty$, according to (\ref{KG1}), we have $\varphi ''\le 0$ near the origin. If the potential is bounded, i.\,e. $\lim\limits_{x\to 0}V=V_0$, where $V_0$ is a negative constant, we multiply both sides of (\ref{KG1}) by $\varphi$ and integrate it to get:
\begin{equation*}
\int_{-\infty}^\infty \varphi '' \varphi dx=\int_{-\infty}^\infty \left[m^2-(E-V)^2\right]\varphi^2dx.
\end{equation*}
After integrating by parts and using $\lim\limits_{x\to\pm\infty}\varphi=0$, the right-hand side of the above expression becomes $-\int_{-\infty}^\infty \left(\varphi ' \right)^2 dx$. The potential $V$ is nondecreasing function so $V_0\le V$, thus we have $\int_{-\infty}^\infty \left[m^2-(E-V_0)^2\right]\varphi^2dx\le \int_{-\infty}^\infty \left[m^2-(E-V)^2\right]\varphi^2dx\le 0$. Therefore $m^2-(E-V_0)^2\le 0$ which means that $\varphi ''\le 0$ near the origin. The wave function $\varphi$ is even and $\varphi(0)=1$, so $\varphi '(0)=0$. Since $\varphi$ is concave near the origin, we conclude that $\varphi '\le 0$ for $x>0$ near zero.

Near positive infinity, according to (\ref{KG1inf}), $\varphi\ge 0$ and $\varphi\in L^2(\mathbb{R})$ thus $\varphi '\le 0$. The function $m^2-(E-V)^2$ is monotone nondecreasing because $\left(m^2-(E-V)^2\right)'\ge 0$ therefore $\varphi ''$ changes sign at most once. If we suppose that $\varphi '>0$ on some interval in $[0,\ \infty)$, then $\varphi ''$ must change sign more than once, which yields a contradiction. Hence $\varphi '\le 0$ on $x\in[0,\ \infty)$ and this corresponds to the lemma's inequality. 

\hfill $\Box$

Before sharpening the comparison theorem, we shall first re-establish the base comparison theorem itself. Therefore we write (\ref{KG1}) for two comparison potentials $V_a$ and $V_b$ and corresponding energy eigenvalues $E_a$ and $E_b$:
\begin{equation}\label{KGda1}
\varphi ''_a=\left[m^2-(E_a-V_a)^2\right]\varphi_a,
\end{equation}
and
\begin{equation}\label{KGdb1}
\varphi ''_b=\left[m^2-(E_b-V_b)^2\right]\varphi_b.
\end{equation}
Then we multiply eqution (\ref{KGda1}) by $\varphi_b$ and subtract (\ref{KGdb1}) multiplied by $\varphi_a$ to get 
\begin{equation}\label{expr}
(\varphi_a'\varphi_b)'-(\varphi_b'\varphi_a)'=W\varphi_a\varphi_b
[(V_b-V_a)-(E_b-E_a)],
\end{equation}
where 
\begin{equation*}
W=E_a+E_b-V_a-V_b.
\end{equation*}
Since $\varphi '(0)=0$ and $\lim\limits_{x\to\infty}\varphi=0$, the left side of (\ref{expr}) after integration by parts is zero and the right side is
\begin{equation}\label{CTd1}
(E_b-E_a)\int_0^\infty W\varphi_a\varphi_bdx 
=\int_0^\infty(V_b-V_a)W\varphi_a\varphi_b dx.
\end{equation}
Here we have arrived at the comparison theorem derived by Hall and Aliyu \cite{HallComp} which states that if $E_a\ge 0$, $E_b\ge 0$, and $V_a\le V_b$ then $E_a\le E_b$. This proof requires that the integrands of (\ref{CTd1}) should have the same signs. Thus $\varphi$ can not have a node, i.e. $\varphi$ corresponds to the ground--state wave function. We note that if we allow negative eigenvalues, then this reasoning would fail since $W$ may not have constant sign.

Now we shall sharpen the comparison theorem and prove: 

\medskip

\noindent{\bf Theorem 1:} ~~{\it If $V$ satisfies $1^\circ$--$\ 4^\circ$, has area, and 
\begin{equation}\label{th_g}
\nu(x)=\int_0^x (V_b(t)-V_a(t)) dt\ge 0, \quad x\in [0,\ \infty),
\end{equation}
then $E_a\le E_b$.} 

\medskip

\noindent{\bf Proof:} Integration by parts of the right side of (\ref{CTd1}) yields
\begin{equation*}
\int_0^\infty(V_b-V_a)W\varphi_a\varphi_b dx
=\left[\varphi_a\varphi_bW\nu(x)\right]_0^\infty - \int_0^\infty \nu(x)(\varphi_a\varphi_bW)'dx,
\end{equation*}
Using $\nu(0)=0$ and $\lim\limits_{x\to\infty}\varphi=0$ in the above expression, we write relation (\ref{CTd1}) as
\begin{equation*}
(E_b-E_a)\int_0^\infty W\varphi_a\varphi_bdx=
- \int_0^\infty \nu(x)(\varphi_a\varphi_bW)'dx.
\end{equation*}
It follows from assumption $3^\circ$ that the function $W$ is nonincreasing, so $W'\le 0$ (a.\,e.) and Lemma 1 gives us $(\varphi_a\varphi_bW)'\le 0$. Thus if $\nu(x)\ge 0$ the above equation implies that $E_a\le E_b$. This completes the proof.

\hfill $\Box$

If the more detailed behaviour of the comparison potentials is known and the potential difference $V_b-V_a$ has finite area, then the simpler sufficient conditions for spectral ordering immediately follow from the above theorem:

\medskip

\noindent{\bf Corollary 1:} ~~{\it If the graphs of the comparison potentials cross over once at $x=x_1$, $V_a\le V_b$ for $x\in [0,\ x_1]$, and
\begin{equation*}
\nu(\infty)=\int_0^\infty (V_b-V_a)dx\ge 0,\quad \text{then}\quad E_a\le E_b. 
\end{equation*}
If the graphs of the comparison potentials cross over twice, at $x=x_1$ and $x=x_2$, $x_1<x_2$, $V_a\le V_b$ for $x\in [0,\ x_1]$, and
\begin{equation*}
\nu(x_2)=\int_0^{x_2} (V_b-V_a) dx\ge 0,\quad \text{then}\quad E_a\le E_b.  
\end{equation*}} 

\medskip

The above corollary gives sufficient conditions for spectral ordering in case of one and two intersections. Using the same reasoning, we can generalize Corollary 1 for the case of $n$ intersections: {\it If the graphs of the comparison potentials cross over at $n$ points, $n=1,\ 2,\ 3,\ \ldots$, $V_a\le V_b$ for $x\in [0,\ x_1]$, 
where $x_1$ is the first intersection point, and the sequence of absolute areas
$\int_{x_i}^{x_{i+1}}|V_b-V_a|dx$, $i=1,\ 2,\ 3,\ \ldots, \ n$, is nonincreasing (if $n$ is odd we suppose $\int_{x_{n-1}}^{x_n}|V_b-V_a|dx\ge\int_{x_n}^\infty|V_b-V_a|dx$), then $\nu\ge 0$ for $x\in[0,\ \infty)$ and $E_a\le E_b$.}

If we know still more details concerning the solution of {\it one} of the two comparison problems, i.\,e. {\it one} of the two wave functions is known either $\varphi_a$ or $\varphi_b$. Then we state a sharper result, namely:

\medskip

\noindent{\bf Theorem 2:} ~~{\it If $V$ satisfies $1^\circ$--$\ 4^\circ$, has $\varphi_j$--weighted area, and 
\begin{equation}\label{th_h}
\mu(x)=\int_0^x (V_b(t)-V_a(t))\varphi_j dt\ge 0, \quad x\in [0,\ \infty),
\end{equation}
then $E_a\le E_b$, where j is either a or b.}

\medskip

\noindent{\bf Proof:} Consider the case $j=b$. Integrating by parts the right side of (\ref{CTd1}) and using  
$\mu(0)=0$ and $\lim\limits_{x\to\infty}\varphi_a=0$, we write relation (\ref{CTd1}) as
\begin{equation*}
(E_b-E_a)\int_0^\infty W\varphi_a\varphi_bdx =
- \int_0^\infty \mu(x)(\varphi_aW)'dx.
\end{equation*}
According to Lemma 1 and assumption $3^\circ$ the derivative $(\varphi_aW)'\le 0$ and if $\mu(x)\ge 0$ then $E_a\le E_b$. Case $j=a$ can be proved in exactly the same way.

\hfill $\Box$

Comparing the above two theorems we note that in Theorem 2 the potential difference $\triangle V = V_b - V_a$ is multiplied by $\varphi_j$, since $\lim\limits_{x\to\infty}\varphi_j=0$, $\triangle V$ can be even larger than in Theorem 1 and still imply the spectral ordering. Following the same path and using the second theorem, we can state a simple sufficient condition for spectral ordering, which is easy to apply in practice:

\medskip

\noindent{\bf Corollary 2:} ~~{\it If the graphs of the comparison potentials cross over once at $x=x_1$, $V_a\le V_b$ for $x\in [0,\ x_1]$, and
\begin{equation*}\label{concl3}
\mu(\infty)=\int_0^\infty (V_b-V_a)\varphi_j dx\ge 0,\quad \text{then}\quad E_a\le E_b,\quad\text{where} \quad j=a \ \text{or} \  b.
\end{equation*}
If the graphs of the comparison potentials cross over twice, at $x=x_1$ and $x=x_2$, $x_1<x_2$, $V_a\le V_b$ for $x\in [0,\ x_1]$, and
\begin{equation*}\label{concl4}
\mu(x_2)=\int_0^{x_2} (V_b-V_a)\varphi_j dx\ge 0,\quad \text{then}\quad E_a\le E_b,\quad\text{where} \quad j=a \ \text{or} \  b.   
\end{equation*}} 

\medskip

As before we can generalize Corollary 2 for the case of $n$ intersections: {\it If the graphs of the comparison potentials cross over at $n$ points, $n=1,\ 2,\ 3,\ \ldots$, $V_a\le V_b$ for $x\in [0,\ x_1]$, 
where $x_1$ is the first intersection point, and the sequence of absolute areas
$\int_{x_i}^{x_{i+1}}|V_b-V_a|\varphi_jdx$, $i=1,\ 2,\ 3,\ \ldots, \ n$, is nonincreasing (if $n$ is odd we suppose $\int_{x_{n-1}}^{x_n}|V_b-V_a|\varphi_jdx\ge\int_{x_n}^\infty|V_b-V_a|\varphi_jdx$), then $\mu\ge 0$ for $r\in[0,\ \infty)$ and $E_a\le E_b$.}

\subsubsection*{An Example}
Let us consider two comparison potentials: the cut--off Coulomb potential $V_a$ \cite{cutoff_2, cutoff_1}, which has a known analytical solution \cite{Barton}, and the sech--squared potential $V_b$ \cite{sechsquared_4, sechsquared_2, sechsquared_1, sechsquared_3}
\begin{equation*}
V_a=-\frac{\alpha}{|x|+a} \qquad \text{and} \qquad 
V_b=-\frac{4\beta}{(e^{bx}+e^{-bx})^2}.
\end{equation*}
Choosing $\alpha=0.74973$, $a=0.5$, $\beta=0.7$, and $b=0.4$, we find that the
\begin{figure}
\centering{\includegraphics[width=8cm,height=14cm,angle=-90]{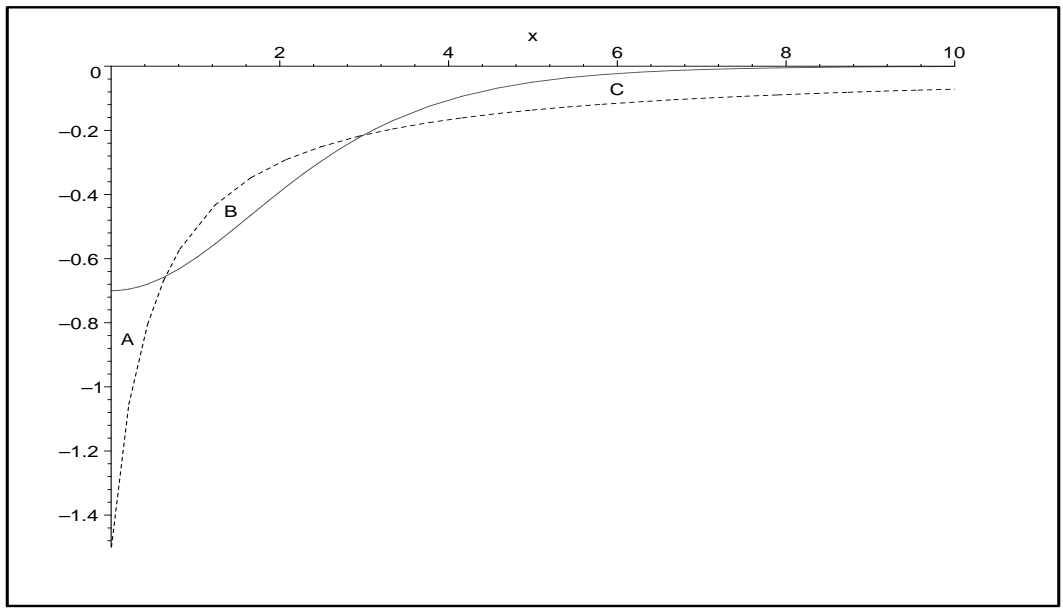}}
\caption{Potentials $V_a=-\cfrac{\alpha}{|x|+a}$ dashed lines and $V_b=-\cfrac{4\beta}{(e^{bx}+e^{-bx})^2}$ full line. Where the following values were applied: $\alpha=0.74973$, $a=0.5$, $\beta=0.7$, and $b=0.4$.}
\end{figure}
graphs of $V_a$ and $V_b$ intersect at $x_1=0.64361$ and $x_2=2.99146$; see Figure 1. Then direct calculation shows
\begin{equation*}
A=\int_0^{x_1} (V_b - V_a)dx=0.17945>B=\int_{x_1}^{x_2} (V_a - V_b)dx=0.17944.
\end{equation*}
Thus, according to Corollary 1, $\nu>0$ which leads to $E_a\le E_b$. We verify this result by calculating accurate numerical eigenvalues: $E_a=0.34055\le E_b=0.48874$.

\section{Sharp comparison theorem in $d>1$ dimensions}
The Klein--Gordon equation in $d$ dimensions is given by \cite{Greiner, Alhaidari}
\begin{equation*}
\triangle_d\Psi({\bf r})=\left[m^2-(E-V(r))^2 \right]\Psi({\bf r}),
\end{equation*}
where natural units $\hbar=c=1$ are used and $E$ is the discrete energy eigenvalue of a spinless particle of mass $m$. The potential function $V(r)$, $r=||{\bf r}||$, is a radially symmetric Lorentz vector potential, and $\triangle_d$ is the $d$--dimensional Laplacian. Thus for $d>1$ the wave function can be written as $\Psi({\bf r})=R(r)Y_{l_{d-1},\ldots,l_1}(\theta_1,\ \theta_2,\ \ldots, \ \theta_{d-1})$, where $R\in L^2(\mathbb{R}^d)$ is a radial function and $Y_{l_{d-1},\ldots,l_1}$ is a normalized hyper--spherical harmonic with eigenvalues $l(l+d-1)$, $l=0,\ 1,\ 2,\ \ldots$ (details can be found in \cite{eval}). Correspondingly, the radial part of the above Klein--Gordon equation may be written as the second--order ordinary differential equation
\begin{equation}\label{KGd_1}
R''+\frac{d-1}{r}R'=\left[m^2-(E-V)^2+\frac{l(l+d-2)}{r^2}\right]R.
\end{equation}
We introduce the reduced radial wave function $\psi$ by $R=r^{-(d-1)/2}\psi$ to get 
\begin{equation}\label{KGd}
\psi ''=\left[m^2-(E-V)^2+\frac{Q}{r^2}\right]\psi,
\end{equation}
where
\begin{equation*}
Q=\frac{1}{4}(2l+d-1)(2l+d-3), \quad l=0, 1, 2, \ldots, \quad \text{and}  \quad d=2, 3, 4, \ldots
\end{equation*}
which is the radial Klein--Gordon equation in $d>1$ dimensions. The reduced radial wave function $\psi$ satisfies $\psi(0)=0$ and $\lim\limits_{r\to\infty}\psi=0$ \cite{Nieto}, and for bound states the normalization condition becomes
\begin{equation*}
\int_0^\infty \psi^2 dr=1.
\end{equation*}
We suppose that the vector potential $V$ (the time component of a four--vector) satisfies
\begin{eqnarray*}
&1^\circ&V\ \text{is nonpositive, i.\,e.} \  V\le 0;\\
\vspace{7mm}
&2^\circ&V\ \text{is attractive, that is to say monotone nondecreasing on} \  [0, \infty), \text{so} \  V'\ge 0;\\
\vspace{7mm}
&3^\circ&V\ \text{vanishes at infinity, thus}\ \lim_{r\to \infty}V=0.
\end{eqnarray*}
Lemma 2 below requires also that in $d=2$ dimensions if the potential is bounded, the function $V$ has to be below $-m$ near the origin  i.\,e. $\lim\limits_{r\to 0^+}V=V_0$, where $V_0$ is negative constant and $V_0<-m$.

Using (\ref{KGd}) and following the same argument as in the one--dimensional case, we can obtain the corresponding equation to that used for the comparison theorem in Ref. \cite{HallComp}. In $d>1$ dimensions for two comparison potentials $V_a$ and $V_b$ we have
\begin{equation}\label{CTd}
(E_b-E_a)\int_0^\infty W\psi_a\psi_bdr
=\int_0^\infty(V_b-V_a)W\psi_a\psi_b dr
\end{equation}
where 
\begin{equation*}
W=E_a+E_b-V_a-V_b.
\end{equation*}
Then it follows from the above expression that if the eigenvalues are positive and $V_a\le V_b$ then $E_a\le E_b$, which is the basic relativistic comparison theorem \cite{HallComp}. We shall sharpen this theorem in the next section for $l=0$ and the following theorem helps to extend our results to the $l>0$ cases. As in the one--dimensional case, we consider only node--free states with nonnegative eigenvalues, so that $0\le E<m$, and without loss of generality we may assume $\psi\ge 0$. 

Similarly to the nonrelativistic case \cite{ddimSch}, we derive a simple relation between discrete eigenvalues for angular momenta $l$ and in dimension $d$. Here  $k=d+2l$.

\medskip

\noindent{\bf Theorem 3:} ~~{\it Let $E_{nl}^d$ be the discrete eigenvalue in $d$ dimensions which corresponds to the radial Klein--Gordon wave function with $n$ nodes, and the angular--momentum subspace labelled by $l$, then 
\begin{equation*}
E_{nl}^d=E_{n0}^{d+2l}.
\end{equation*}} 

\medskip

\noindent{\bf Proof:} We rewrite the radial eigenequation (\ref{KGd}) in the following form
\begin{equation*}
\psi ''=\left[m^2-(E_{nl}^d-V)^2+\frac{(k-1)(k-3)}{r^2}\right]\psi,
\end{equation*}
where $k=d+2l$ and $E_{nl}^d$ corresponds to $\psi$. The above equation can also be seen as the radial Klein--Gordon equation in $k$ dimensions and $l=0$. Thus $E_{nl}^d=E_{n0}^k=E_{n0}^{d+2l}$. 

\hfill $\Box$

Because of the preceding theorem all the comparison results which we derive in $d>1$ dimensions for $n=0$ and $l=0$, can be then applied to the family of equivalent problems in $d'$ dimensions for $n=0$, $l>0$, and $d'=d-2l\ge 2$.  

To sharpen the comparison theorem, which follows from relation (\ref{CTd}), we need to know more about the behaviour of the wave function:

\medskip

\noindent{\bf Lemma 2:} ~~{\it The Klein--Gordon $s$--state energy eigenfunction $\psi$ satisfies 
\begin{equation}\label{claim2}
\left(\frac{\psi}{r^{\frac{d-1}{2}}}\right)'\le 0, \quad r\in[0,\ \infty).
\end{equation}} 

\medskip

\noindent{\bf Proof:} Equation (\ref{KGd_1})  for the case $l=0$ becomes
\begin{equation}\label{eq1}
R''+\frac{d-1}{r}R'=\left[m^2-(E-V)^2\right]R.
\end{equation}
We then rewrite it in the following way:
\begin{equation}\label{eq2}
\frac{1}{t^{d-1}}\frac{\partial}{\partial t}\left(t^{d-1}\frac{\partial}{\partial t}\right) R(t)=F(t)R(t),
\end{equation}
where $F=m^2-(E-V)^2$.
It follows from (\ref{eq2}) that
\begin{equation}\label{eq3}
R'=r^{-(d-1)}\int_0^rF(t)R(t)t^{d-1}dt.
\end{equation}
Since $R=r^{-(d-1)/2}\psi$ we have to prove that $R'\le 0$, where $R'$ is given by $(\ref{eq3})$. 
Consider the function $F$: we know that $\lim\limits_{r\to\infty}V=0$; thus $\lim\limits_{r\to\infty}F=\lim\limits_{r\to\infty}[m^2-(E-V)^2]=m^2-E^2$. The eigenvalue $E$ is such that $0\le E<m$. Therefore $m^2-E^2>0$ which means that the function $F$ is positive near infinity. 

Now we prove that $F$ must change sign for $r\in[0,\ \infty)$. Equation (\ref{KGd}) for $l=0$ becomes
\begin{equation}\label{eq4}
\psi ''=\left[F+\frac{(d-1)(d-3)}{4r^2}\right]\psi.
\end{equation}
Then, since, without loss of generality, we consider nonnegative $\psi$, the assumption that $F>0$ on $[0,\ \infty)$ and equation (\ref{eq4}) yield $\psi ''\ge 0$ in $d\ge 3$ dimensions\footnote{In $d=2$ dimensions if the potential function $V$ is unbounded, i. e. $\lim\limits_{r\to 0}V=-\infty$, $\lim\limits_{r\to 0}F=-\infty$ so $F<0$ for small $r$. If potential is bounded i. e. $\lim\limits_{r\to 0}V=V_0$, where $V_0$ is negative constant, the condition $V_0<-m$ ensures that $F< 0$ near the zero.}. But this contradicts the facts that $\psi\ge 0$, $\psi(0)=0$, and $\lim\limits_{r\to\infty}\psi=0$. Therefore $F$ changes sign. The function $F$ is nondecreasing because $F'=2(E-V)V'\ge 0$ since $V$ satisfies $1^\circ$ and $2^\circ$. Thus we finally conclude that $F$ changes sign from negative to positive and therefore there exists one point $\hat{r}$ such that $F(\hat{r})=0$. 

We now return to equation (\ref{eq3}): since $F\le 0$ for $r\le \hat{r}$, $F\ge 0$ for $r\ge \hat{r}$, and $R\ge 0$ on $[0,\ \infty)$, we have $\int_0^{\hat{r}}F(t)R(t)t^{d-1}dt\le 0$ and $\int_{\hat{r}}^\infty F(t)R(t)t^{d-1}dt\ge 0$. We suppose that there exists a point $\bar{r}$ such that $\hat{r}<\bar{r}<\infty$ and 
\begin{equation*}
-\int_0^{\hat{r}} F(t)R(t)t^{d-1}dt=\int_{\hat{r}}^{\bar{r}} F(t)R(t)t^{d-1}dt.
\end{equation*} 
If such a point $\bar{r}$ exists then $\int_0^\infty F(t)R(t)t^{d-1}dt>0$ which leads to $R'>0$ on $[0,\ \infty)$. We know that $R\ge 0$ and $R\in L^2(\mathbb{R}^d)$, which contradicts $R'>0$. Therefore we conclude that a point $\bar{r}$ does not exist. Thus $R'\le 0$.

\hfill $\Box$

Now we state and prove a sharp comparison theorem in $d$ dimensions:

\medskip

\noindent{\bf Theorem 4:} ~~{\it If $V$  satisfies $1^\circ$--$\ 3^\circ$, 
has $t^{d-1}$--weighted area, and   
\begin{equation}\label{th4}
\rho(r)=\int_0^r (V_b(t)-V_a(t)) t^{d-1} dt\ge 0, \quad r\in[0,\ \infty).
\end{equation}
then $E_a\le E_b$.} 

\medskip

\noindent{\bf Proof:} After integration by parts the right side of (\ref{CTd}) becomes 
\begin{equation*}
\int_0^\infty(V_b-V_a)W\psi_a\psi_b dr
=\left.\rho(r)\frac{W\psi_a\psi_b}{r^{d-1}}\right|_0^\infty - 
\int_0^\infty \rho(r) \left(\frac{W\psi_a\psi_b}{r^{d-1}}\right)' dr,
\end{equation*}
Since $\rho(0)=0$ and $\lim\limits_{r\to\infty}\psi=0$ relation (\ref{CTd}) becomes
\begin{equation*}
(E_b-E_a)\int_0^\infty W\psi_a\psi_bdr
= -\int_0^\infty \rho(r) \left(\frac{W\psi_a\psi_b}{r^{d-1}}\right)'.
\end{equation*}
Since $W$ is a nonnegative nonincreasing function, according to Lemma 2, the derivative $\left(\cfrac{W\psi_a\psi_b}{r^{d-1}}\right)'$ is nonpositive. The theorem's assumption (\ref{th4}) then implies $E_a\le E_b$.

\hfill $\Box$

As in one dimension, if more detailed behaviour of the comparison potentials is known and the weighted (by $r^{d-1}$) potential difference $\triangle V$ has finite area, we can state simpler sufficient conditions for spectral ordering:

\medskip

\noindent{\bf Corollary 4:} ~~{\it If the graphs of the comparison potentials cross over once at $r=r_1$, $V_a\le V_b$ for $r\in [0,\ r_1]$, and
\begin{equation*}
\rho(\infty)=\int_0^\infty (V_b-V_a)r^{d-1} dr\ge 0,\quad \text{then}\quad E_a\le E_b. 
\end{equation*}
If the graphs of the comparison potentials cross over twice, at $r=r_1$ and $r=r_2$, $r_1<r_2$, $V_a\le V_b$ for $r\in [0,\ r_1]$, and
\begin{equation*}
\rho(r_2)=\int_0^{r_2} (V_b-V_a)r^{d-1} dr\ge 0, \quad \text{then}\quad E_a\le E_b.  
\end{equation*}} 

\medskip

As in one dimensional case we can generalize Corollary 4 up to $n$ 
intersections: {\it If the graphs of the comparison potentials cross over at $n$ points, $n=1,\ 2,\ 3,\ \ldots$, $V_a\le V_b$ for $r\in [0,\ r_1]$, 
where $r_1$ is the first intersection point, and the sequence of absolute areas
$\int_{r_i}^{r_{i+1}}|V_b-V_a|r^{d-1}dr$, $i=1,\ 2,\ 3,\ \ldots, \ n$, is nonincreasing (if $n$ is odd we suppose $\int_{r_{n-1}}^{r_n}|V_b-V_a|r^{d-1}dr\ge\int_{r_n}^\infty|V_b-V_a|r^{d-1}dr$), then $\rho\ge 0$ for $r\in[0,\ \infty)$ and $E_a\le E_b$.}

\medskip

\noindent{\bf Theorem 5:} ~~{\it If $V$  satisfies $1^\circ$--$\ 3^\circ$, 
has $\psi_jt^{\frac{d-1}{2}}$--weighted area, and 
\begin{equation}\label{th_k}
\eta(r)=\int_0^r (V_b(t)-V_a(t))\psi_j(t) t^{\frac{d-1}{2}} dt\ge 0, 
\quad r\in[0,\ \infty),
\end{equation}
then $E_a\le E_b$, where $j$ is either $a$ or $b$.}

\medskip

\noindent{\bf Proof:} Consider the case $j=b$. We integrate the right side of (\ref{CTd}) by parts and use $\eta(0)=0$ and $\lim\limits_{r\to\infty}\psi_a=0$ to obtain
\begin{equation*}
(E_b-E_a)\int_0^\infty W\psi_a\psi_bdr
= - \int_0^\infty \eta(r) \left(\frac{W\psi_a}{r^{\frac{d-1}{2}}}\right)' dr.
\end{equation*}
$W$ is nonnegative nonincreasing function, therefore it follows from Lemma 2 that $\left(\cfrac{W\psi_a}{r^{\frac{d-1}{2}}}\right)'\le 0$ and if $\eta(r)\le 0$ the theorem's statement immediately follows. Case $j=a$ can be proved in exactly the same way.

\hfill $\Box$

We note that the above theorem for $d=3$ can be seen and proved also as a sharpened comparison theorem corresponding to the first odd exited state in one spatial dimension. As in the one--dimensional case, if we know more detailed behaviuor of the comparison potentials and one of the wave functions, we can state simpler sufficient conditions:

\medskip

\noindent{\bf Corollary 5:} ~~{\it If the graphs of the comparison potentials cross over once at $r=r_1$, $V_a\le V_b$ for $r\in [0,\ r_1]$, and
\begin{equation*}
\eta(\infty)=\int_0^\infty (V_b-V_a)\psi_j r^{\frac{d-1}{2}} dr\ge 0,\quad \text{then}\quad E_a\le E_b,\quad\text{where} \quad j=a \ \text{or} \  b.
\end{equation*}
If the graphs of the comparison potentials cross over twice, at $r=r_1$ and $r=r_2$, $r_1<r_2$, $V_a\le V_b$ for $r\in [0,\ r_1]$, and
\begin{equation*}
\eta(r_2)=\int_0^{r_2} (V_b-V_a)\psi_j r^{\frac{d-1}{2}} dr\ge 0,\quad \text{then}\quad E_a\le E_b,\quad\text{where} \quad j=a \ \text{or} \  b. 
\end{equation*}} 

\medskip

As in the one dimensional case we can generalize Corollary 5 to allow $n$ 
intersections: {\it If the graphs of the comparison potentials cross over at $n$ points, $n=1,\ 2,\ 3,\ \ldots$, $V_a\le V_b$ for $r\in [0,\ r_1]$, 
where $r_1$ is the first intersection point, and the sequence of absolute areas
$\int_{r_i}^{r_{i+1}}|V_b-V_a|\psi_j r^{\frac{d-1}{2}}dr$, $i=1,\ 2,\ 3,\ \ldots, \ n$, is nonincreasing (if $n$ is odd we suppose $\int_{r_{n-1}}^{r_n}|V_b-V_a|\psi_j r^{\frac{d-1}{2}}dr\ge\int_{r_n}^\infty|V_b-V_a|\psi_j r^{\frac{d-1}{2}}dr$), then $\eta\ge 0$ for $r\in[0,\ \infty)$ and $E_a\le E_b$.}

\subsubsection*{An Example}
Here we demonstrate the extension of Corollary 4 for the case of many intersections. We consider the following comparison potentials
\begin{equation*}
V_a=-\frac{\alpha}{re^{ar}}\left(1+\frac{v\sin(sr)}{re^{\kappa r}}\right) \qquad \text{and} \qquad 
V_b=-\frac{\beta}{re^{br}},
\end{equation*}
which satisfy $1^\circ-3^\circ$ and with the following choice of parameters: $\alpha=\beta=0.2$, $a=\kappa=b=0.2$, $v=0.4$, $s=3$, and $d=3$, intersect at many points; see Figure 2, left graph. Then the integral (\ref{th4}) becomes
\begin{equation*}
\rho(r)=\int_0^\infty (V_b-V_a)r^{2}dt=\alpha v\int_0^\infty\frac{\sin(sr)}{e^{2ar}}dt. 
\end{equation*}
The graph of the integrand is plotted in Figure 3.
\begin{figure}[ht]
\begin{center}$
\begin{array}{cc}
\includegraphics[width=2.5in, angle=-90]{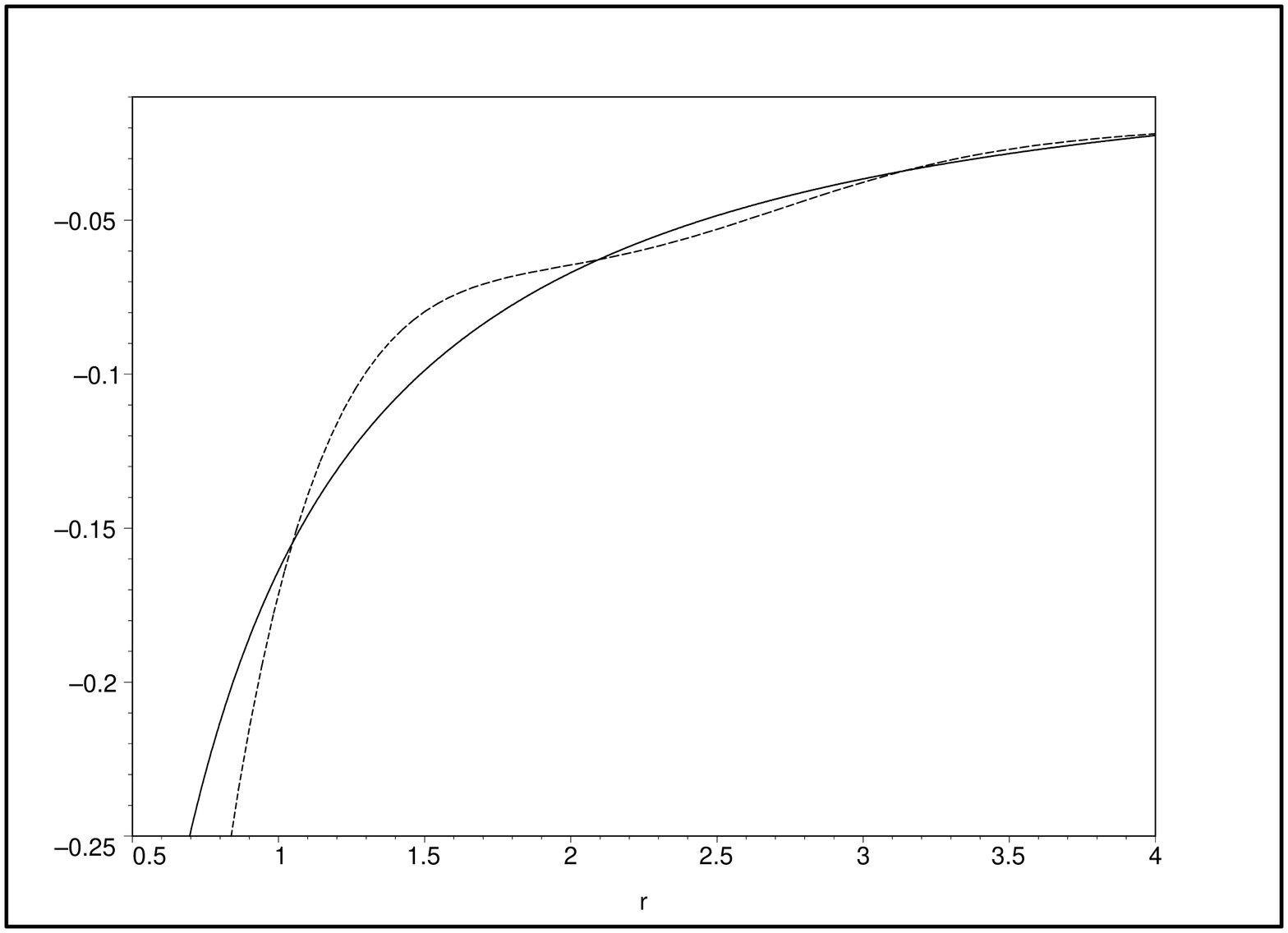} & 
\includegraphics[width=2.5in, angle=-90]{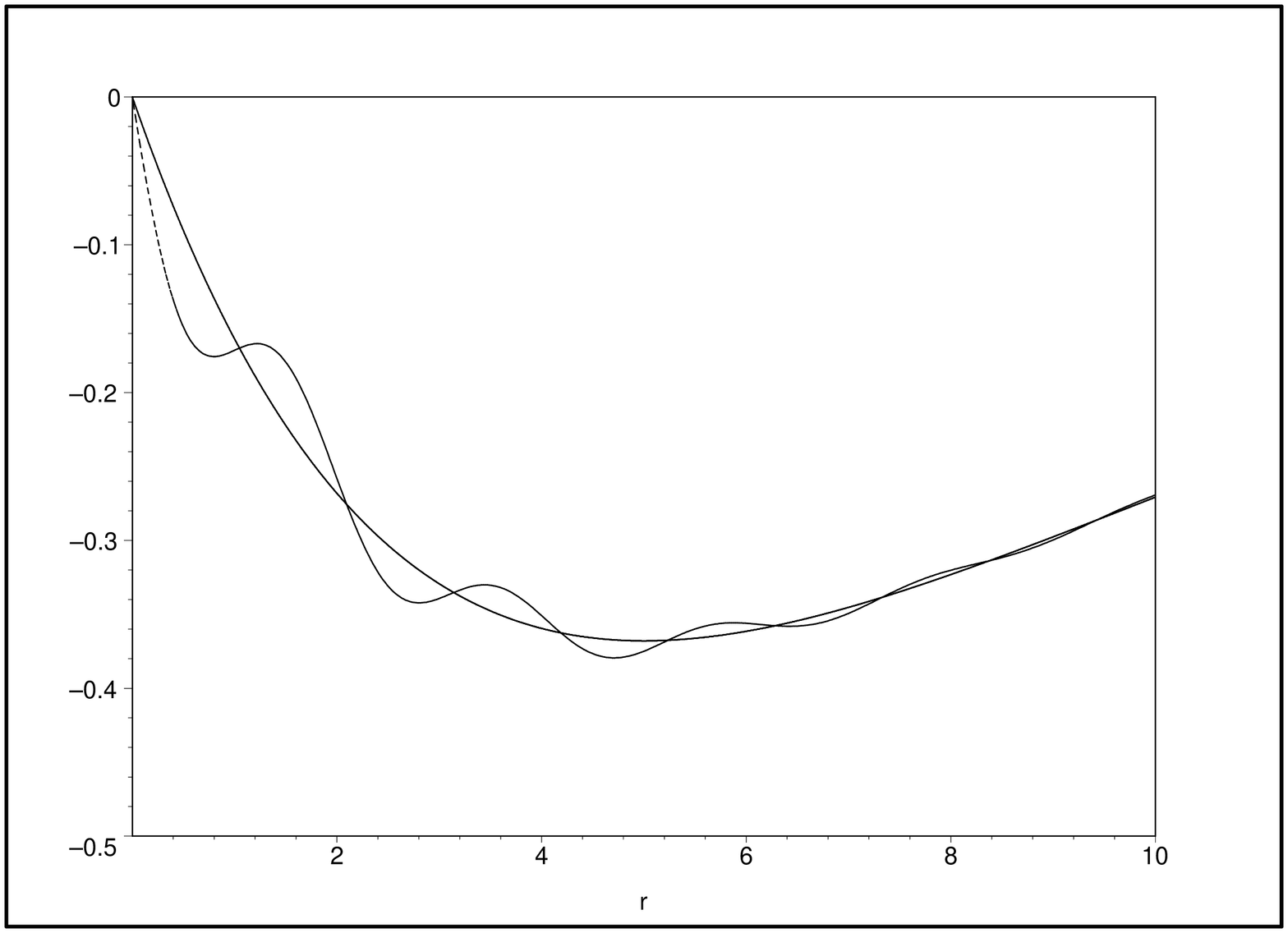}
\end{array}$
\end{center}
\caption{Left graph: potential $V_a=-\cfrac{\alpha}{re^{ar}}\left(1+\cfrac{v\sin(sr)}{re^{\kappa r}}\right)$ dashed lines and $V_b=-\cfrac{\beta}{re^{br}}$ full line. Right graph: function $V_ar^2$ dashed lines and $V_br^2$ full line. Where the following values were applied: $\alpha=\beta=0.2$, $a=\kappa=b=0.2$, $v=0.4$, $s=3$.}
\end{figure} 
\begin{figure}
\centering{\includegraphics[height=15cm,width=6cm,angle=-90]{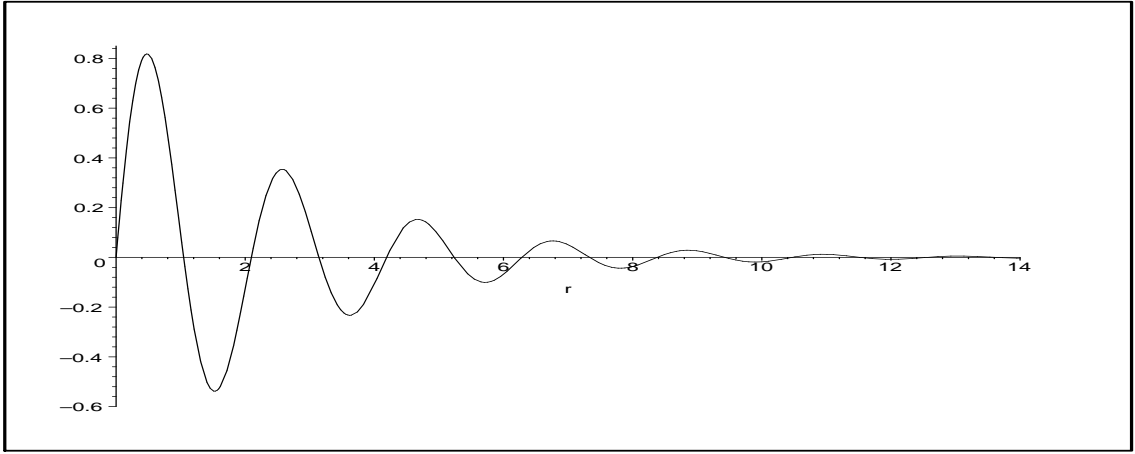}}
\caption{The integrand $I=\cfrac{\sin(sr)}{e^{2ar}}$, where $s=3$ and $a=0.2$.}
\end{figure}
The factor $\sin (sr)$ is a periodic function and $|\sin sx|=|\sin sy|$ where $x\in\left[\cfrac{\pi k}{s},\ \cfrac{\pi(k+1)}{s}\right]$, and $y=x+\cfrac{\pi}{s}$, $k=0,\ 1,\ 2,\ \ldots$. Since $\lim\limits_{r\to\infty}e^{2ar}=\infty$, the absolute value of the integrand areas do not increase, i.\,e.
\begin{equation*}
\int_{\frac{\pi k}{s}}^{\frac{\pi (k+1)}{s}} \frac{|\sin sr|}{e^{2ar}}dr >\int_{\frac{\pi (k+1)}{s}}^{\frac{\pi (k+2)}{s}} \frac{|\sin sr|}{e^{2ar}}dr.
\end{equation*}
Thus  
\begin{equation*}
\int_0^\infty (V_b-V_a)r^{2}dt\ge 0
\end{equation*}
and $\rho>0$, so according to Theorem 4, we conclude $E_a\le E_b$.  
Choosing $m=1$, we verify this result by calculating accurate numerical eigenvalues which are $E_a=0.996204\le E_b=0.999353$.

\section{Sharpened comparison theorems in the presence of scalar potential $S$ in $d>1$ dimensions.}
The radial Klein--Gordon equation (\ref{KGd_1}) in the presence of the scalar potential $S$ is 
\begin{equation}\label{KGd_3}
\triangle R=\left[(m+S)^2-(E-V)^2+\frac{l(l+d-2)}{r^2}\right]R,
\end{equation}
where $\triangle=\cfrac{1}{r^{d-1}}\cfrac{\partial}{\partial r}\left(r^{d-1}\cfrac{\partial}{\partial r}\right)$. Taking $R=r^{-(d-1)/2}\psi$ we get 
\begin{equation}\label{KGVS_1}
\psi '' =\left[(m+S)^2-(E-V)^2+\frac{Q}{r^2}\right]\psi,
\end{equation}
where
\begin{equation*}
Q=\frac{1}{4}(2l+d-1)(2l+d-3), \quad l=0, 1, 2, \ldots,  \quad d=2, 3, 4, \ldots.
\end{equation*}
The reduced radial wave function $\psi$ is zero at the origin and vanishes at infinity, so $\psi(0)=\lim\limits_{r\to\infty}\psi=0$ and satisfies the normalization condition: 
\begin{equation*}
\int_0^\infty\psi^2dr=1. 
\end{equation*}
For different vector $V_i$ and scalar potentials $S_i$, $i=a$ or $b$, the above equation yields
\begin{equation}\label{fir}
\psi_a''=\left[(m+S_a)^2-(E_a-V_a)^2+\frac{Q}{r^2} \right]\psi_a
\end{equation}
and
\begin{equation}\label{sec}
\psi_b''=\left[(m+S_b)^2-(E_b-V_b)^2+\frac{Q}{r^2}\right]\psi_b.
\end{equation}
By considering $\int_0^\infty [(\ref{fir})\psi_b - (\ref{sec})\psi_a]dr$, we obtain first
\begin{equation*}
\int_0^\infty \left[ (E_a-V_a)^2 - (m+S_a)^2 - (E_b-V_b)^2 + (m+S_b)^2 \right]\psi_a\psi_b dr=0,
\end{equation*}
and then
\begin{equation}\label{rel}
(E_b - E_a) \int_0^\infty W\psi_a \psi_bdr=
\int_0^\infty \left[ (V_b - V_a)W + 2m(S_b - S_a) + (S_b^2 - S_a^2)  \right]\psi_a\psi_bdr 
\end{equation}
where 
\begin{equation*}
W=E_a+E_b-V_a-V_b.
\end{equation*}
Here we have arrived at the comparison theorem derived by Luo {\it et al} \cite{Cui}, which states that if $V_a\le V_b\le 0$, $|S_a|\le S_b$, then for positive $E_a$ and $E_b$ we have $E_a\le E_b$.

\subsection{Spin and pseudo--spin symmetry}
As it was treated in \cite{monoton3} we can merge spin symmetry, $S=V$, and pseudo--spin symmetry, $S=-V$, by using the parameter $s$, which is $1$ if $S=V$ and $-1$ if $S=-V$, so $S=sV$. Then equation (\ref{KGVS_1}) for $l=0$ and relation (\ref{rel}) in case $S=sV$ become respectively 
\begin{equation}\label{VS_1}
\psi '' =\left[m^2-E^2+2V(E+sm)+\frac{(d-1)(d-3)}{4r^2}\right]\psi.
\end{equation}
and
\begin{equation}\label{rel2}
(E_b - E_a) \int_0^\infty W\psi_a \psi_bdr=
w\int_0^\infty(V_b - V_a)\psi_a\psi_bdr, 
\end{equation}
where 
\begin{equation*}
w=E_a+E_b+2sm.
\end{equation*}
Equation (\ref{VS_1}) is a Schr\"{o}dinger--like equation. Thus, according to the spectral properties of the Schr\"{o}dinger operator \cite{Reed}, we consider the energy eigenvalue $E$ such that $0\le sE<m$ for the following class of potentials in the $S=sV$ case:
\begin{eqnarray*}
&4^\circ&V\ \text{is nonpositive if $s=1$ and nonnegative if $s=-1$, i.\,e.} \  sV\le 0;\\
\vspace{7mm}
&5^\circ&V\ \text{is monotone on} \  [0, \infty), \text{so} \  sV'\ge 0;\\
\vspace{7mm}
&6^\circ&V\ \text{vanishes at infinity, thus}\ \lim_{r\to \infty}V=0.
\end{eqnarray*}

We note that in $d=2$ case, if the potential function is bounded at the origin, i.\ e. $\lim\limits_{r\to 0^+}V=V_0$, where $V_0$ is a constant, then $sV_0<-m/2$ and the following lemma can be proved:

\medskip

\noindent{\bf Lemma 3:} ~~{\it The Klein--Gordon $s$--state energy eigenfunction $\psi$, in the case $S=sV$, satisfies
\begin{equation*}
\left(\frac{\psi}{r^{\frac{d-1}{2}}}\right)'\le 0, \quad r\in[0,\ \infty).
\end{equation*}} 

\medskip

\noindent{\bf Proof:} The proof is similar to the second lemma's proof, i.\,e. using equation (\ref{KGd_3}) one finds
\begin{equation*}
R'=r^{-(d-1)}\int_0^r F(t)R(t)t^{d-1}dt,
\end{equation*}
where $F=2V(E+sm)+m^2-E^2$. Then, it can be shown that there is one point $\hat{r}>0$ satisfying $F(\hat{r})=0$. Finally, by contradiction, one can obtain 
$R'=\left(\cfrac{\psi}{r^{\frac{d-1}{2}}}\right)'\le 0$ for $r\in[0,\ \infty)$, which result completes the proof of Lemma 3. 

\hfill $\Box$

Then, using relation (\ref{rel2}) and Lemma 3, the below theorems and following corollaries can be established:

\medskip

\noindent{\bf Theorem 6:} ~~{\it If $V$  satisfies $4^\circ$--$\ 6^\circ$, 
has $t^{d-1}$--weighted area, $S=sV$, and   
\begin{equation*}
\xi(r)=\int_0^r (V_b(t)-V_a(t)) t^{d-1} dt\ge 0, \quad r\in[0,\ \infty).
\end{equation*}
then $E_a\le E_b$.} 

\medskip

\noindent{\bf Corollary 6:} ~~{\it If the graphs of the comparison potentials cross over once at $r=r_1$, $V_a\le V_b$ for $r\in [0,\ r_1]$, and
\begin{equation*}
\xi(\infty)=\int_0^\infty (V_b-V_a)r^{d-1} dr\ge 0,\quad \text{then}\quad E_a\le E_b. 
\end{equation*}
If the graphs of the comparison potentials cross over twice, at $r=r_1$ and $r=r_2$, $r_1<r_2$, $V_a\le V_b$ for $r\in [0,\ r_1]$, and
\begin{equation*}
\xi(r_2)=\int_0^{r_2} (V_b-V_a)r^{d-1} dr\ge 0, \quad \text{then}\quad E_a\le E_b.  
\end{equation*}} 

\medskip

\noindent{\bf Theorem 7:} ~~{\it If $V$  satisfies $4^\circ$--$\ 6^\circ$, 
has $\psi_jt^{\frac{d-1}{2}}$--weighted area, $S=sV$, and 
\begin{equation*}
\zeta(r)\int_0^r (V_b(t)-V_a(t))\psi_j(t) t^{\frac{d-1}{2}} dt\ge 0, 
\quad r\in[0,\ \infty),
\end{equation*}
then $E_a\le E_b$, where $j$ is either $a$ or $b$.}

\medskip

\noindent{\bf Corollary 7:} ~~{\it If the graphs of the comparison potentials cross over once at $r=r_1$, $V_a\le V_b$ for $r\in [0,\ r_1]$, and
\begin{equation*}
\zeta(\infty)=\int_0^\infty (V_b-V_a)\psi_j r^{\frac{d-1}{2}} dr\ge 0,\quad \text{then}\quad E_a\le E_b,\quad\text{where} \quad j=a \ \text{or} \  b.
\end{equation*}
If the graphs of the comparison potentials cross over twice, at $r=r_1$ and $r=r_2$, $r_1<r_2$, $V_a\le V_b$ for $r\in [0,\ r_1]$, and
\begin{equation*}
\zeta(r_2)=\int_0^{r_2} (V_b-V_a)\psi_j r^{\frac{d-1}{2}} dr\ge 0,\quad \text{then}\quad E_a\le E_b,\quad\text{where} \quad j=a \ \text{or} \  b. 
\end{equation*}} 

\medskip

As before we can generalize corollaries 6 and 7 up to the case of $n$ intersections.

\subsubsection*{An Example}
As an example we consider Theorem 7, in particular first part of Corollary 7 with one intersection point. We take the Yukawa potential $V_a$ \cite{Yuk} and the Hulth\'{e}n potential $V_b$ \cite{Hult_1, Hult_2} in the following form
\begin{equation*}
V_a=-\frac{\alpha}{re^{ra}} \qquad \text{and} \qquad 
V_b=-\frac{\beta}{e^{r/b} - 1}.
\end{equation*}
For $\alpha=0.9$, $a=0.5$, $\beta=0.39$, and $b=1.88$ the potentials cross over once at $r_1=0.90765$; see Figure 4, left graph.
\begin{figure}[ht]
\begin{center}$
\begin{array}{cc}
\includegraphics[width=2.5in, angle=-90]{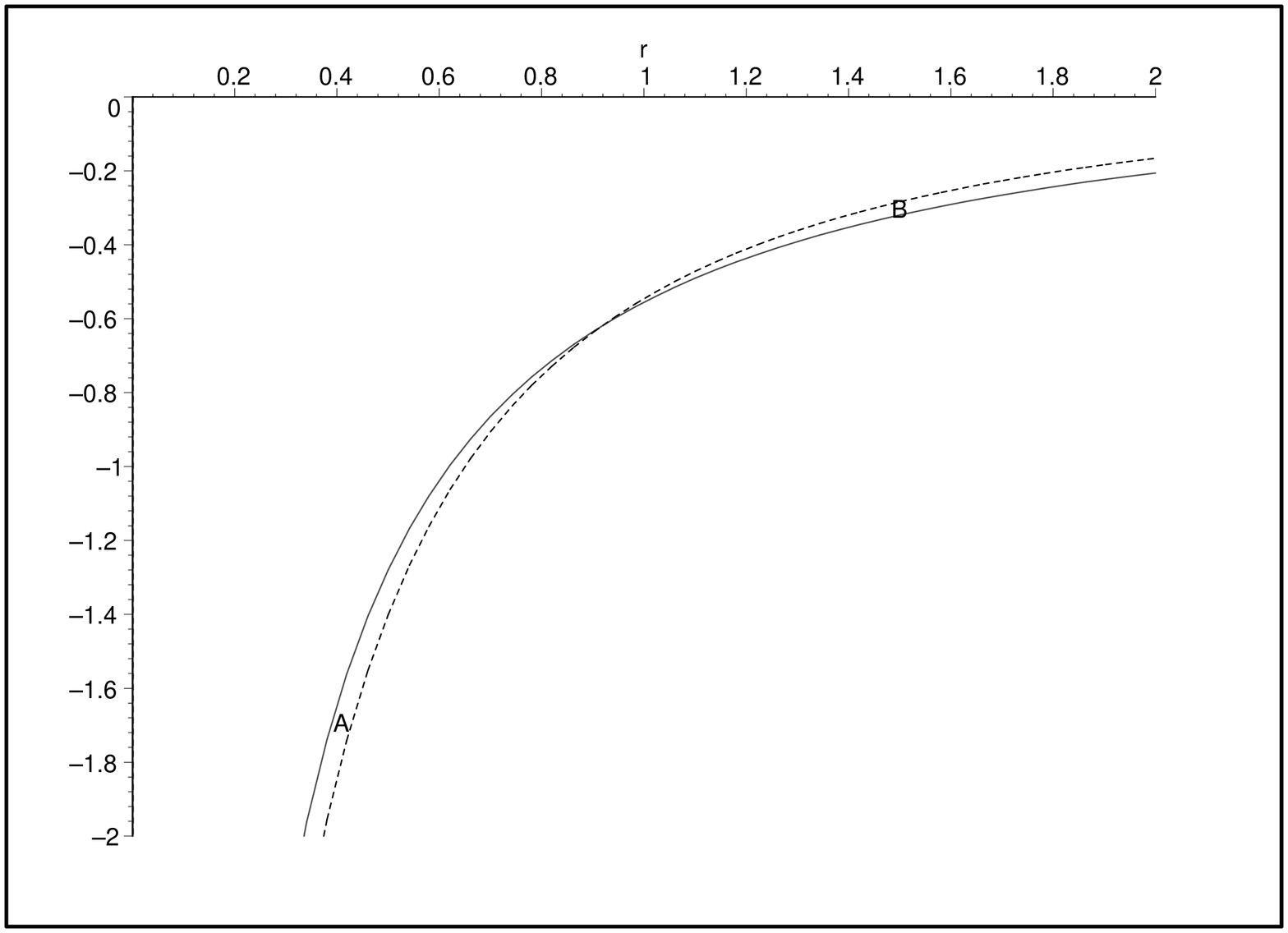} & 
\includegraphics[width=2.5in, angle=-90]{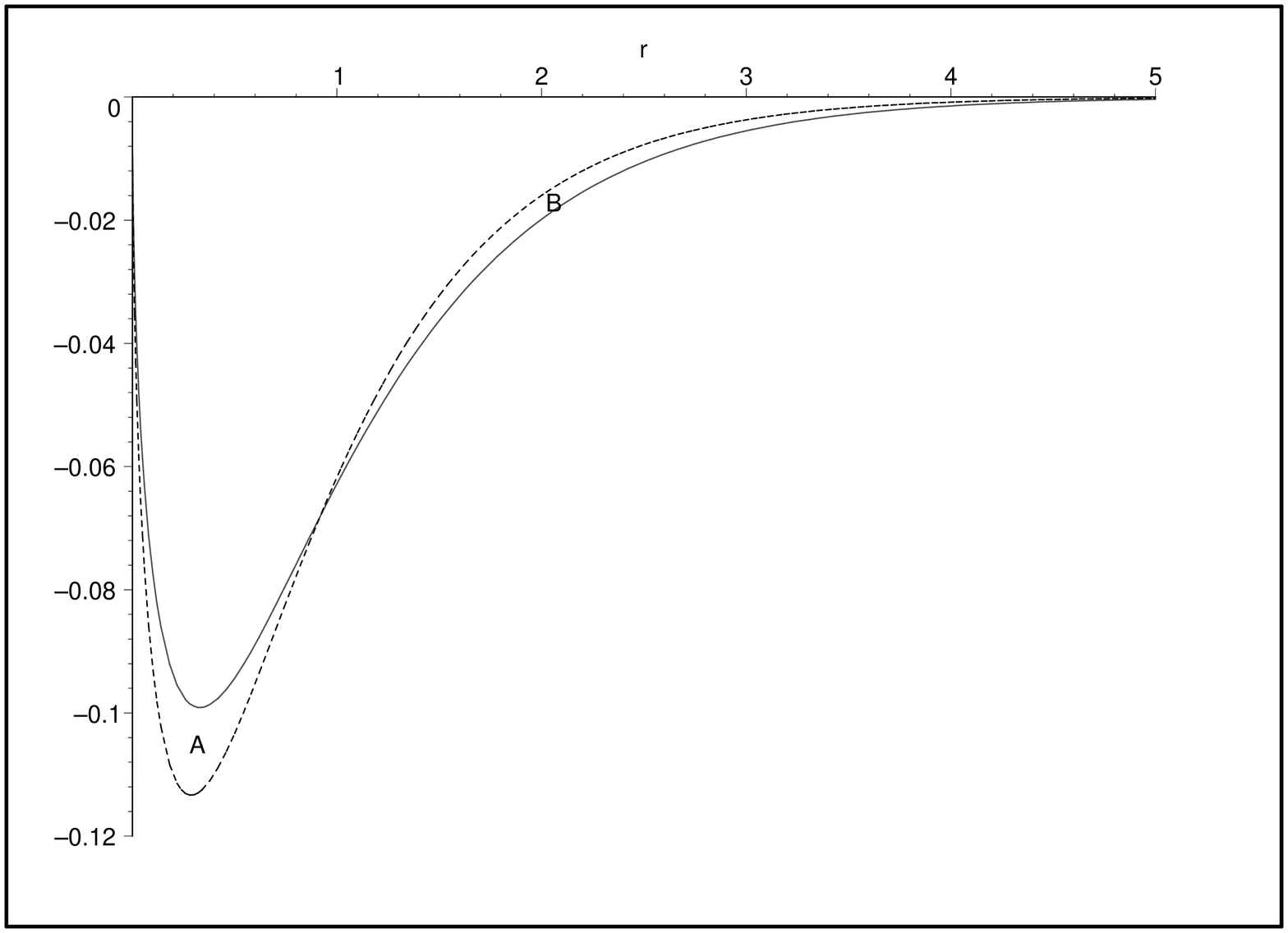}
\end{array}$
\end{center}
\caption{Left graph: potential $V_a=-\cfrac{\alpha}{re^{ra}}$ dashed lines and $V_b=-\cfrac{\beta}{e^{r/b} - 1}$ full line. Right graph: function $V_a\psi_br$ dashed lines and $V_b\psi_br$ full line. Where the following values were applied: $\alpha=0.9$, $a=0.5$, $\beta=0.39$, and $b=1.88$.}
\end{figure}
An article by Dominguez--Adame \cite{Domingues} provides us with the ground--state wave function in $d=3$ dimensions\\ $\psi_b=e^{-r\sqrt{m^2-E_b^2}}\left(1-e^{-r/b}\right)^{3/2}/r$ and the analytical solutions $\sqrt{m^2 - E_b^2}=\beta b (E_b+m)-\cfrac{1}{2b}$ for the case $V=S$. We then calculate the $\psi_b$--weighted areas $A$ and $B$
\begin{equation*}
A=\int_0^{r_1} (V_b-V_a)\psi_brdr=8.15524\times 10^{-3} 
\end{equation*}
and
\begin{equation*}
B=\int_{r_1}^\infty (V_a-V_b)\psi_brdr=8.155239\times 10^{-3}. 
\end{equation*}
Thus $\zeta(r)>0$ and it follows from Corollary 7 that $E_a\le E_b$, which is verified by the accurate numerical eigenvalues $E_a=0.48678\le E_b=0.52421$. 


If potential $V_a$ is bounded and $V_b$ is not then it is not possible to fulfil the corollary condition, i.e. $V_a\le V_b$ for $r\in [0,\ r_1]$. In such a case we may adjust the first part of Corollary 7 in the following way: {\it If the potentials cross over once, say at $r_1$, $V_a\le V_b$ for $r\in [r_1,\ \infty)$, and
\begin{equation*}\label{concl9}
\zeta(\infty)=\int_0^\infty (V_b-V_a)\psi_jr^{\frac{d-1}{2}} dr\ge 0,\quad \text{then}\quad E_a\le E_b,\quad\text{where} \quad j=a \ \text{or} \  b.
\end{equation*}} 
Let us take the soft--core potential $V_a$ and the Hulth\'{e}n potential $V_b$ in $d=3$ dimensions
\begin{equation*}
V_a=-\frac{\alpha}{(r^q + a^q)^{1/q}} \qquad \text{and} \qquad 
V_b=-\frac{\beta}{e^{r/b} - 1},
\end{equation*}
The comparison potentials have one intersection point at $r_1=0.59876$ if $\alpha=0.8$, $a=0.5$, $q=5$, $\beta=1.39$, and $b=0.8$; see Figure 5, left graph.
\begin{figure}[ht]
\begin{center}$
\begin{array}{cc}
\includegraphics[width=2.5in, angle=-90]{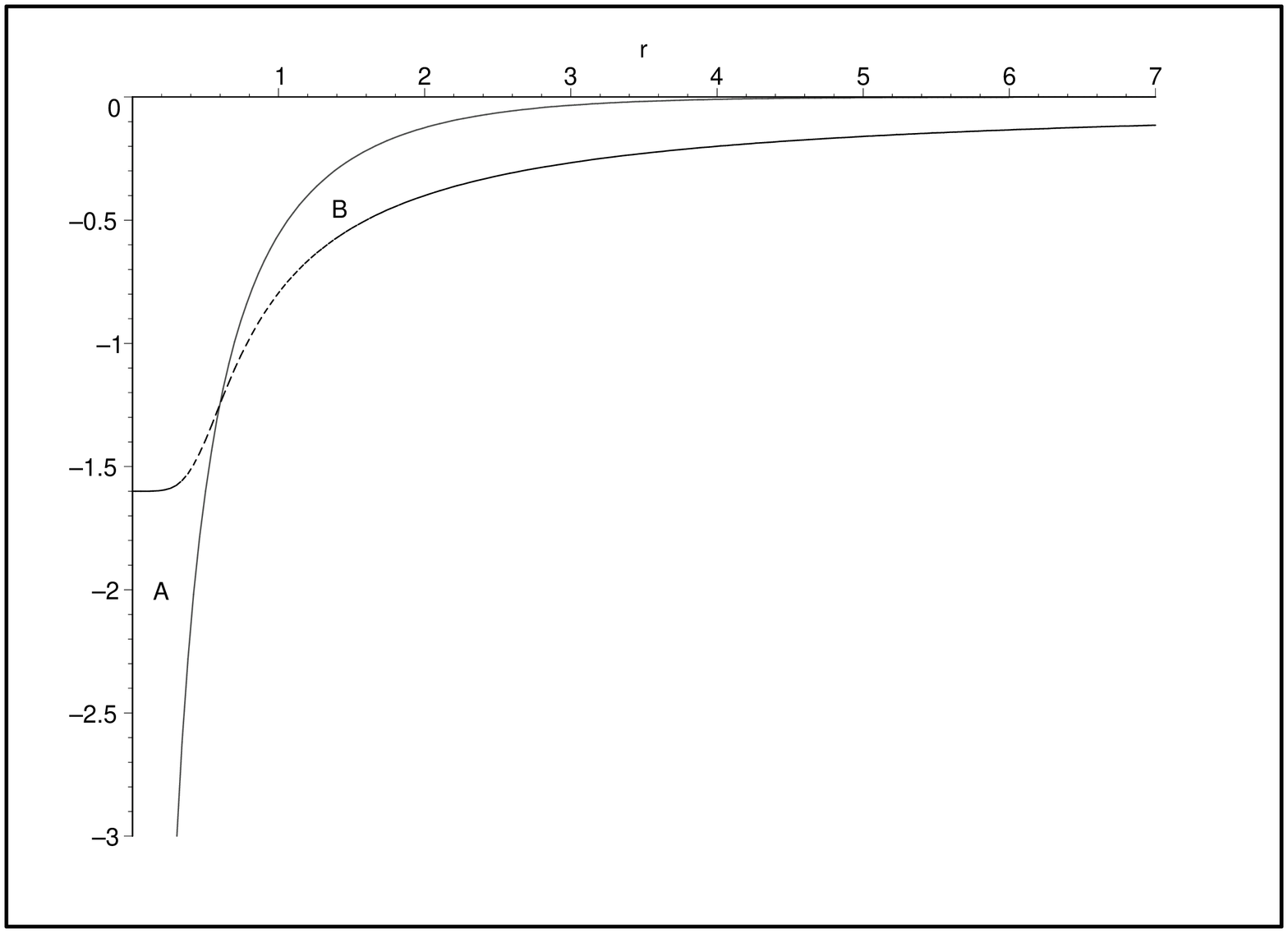} & 
\includegraphics[width=2.5in, angle=-90]{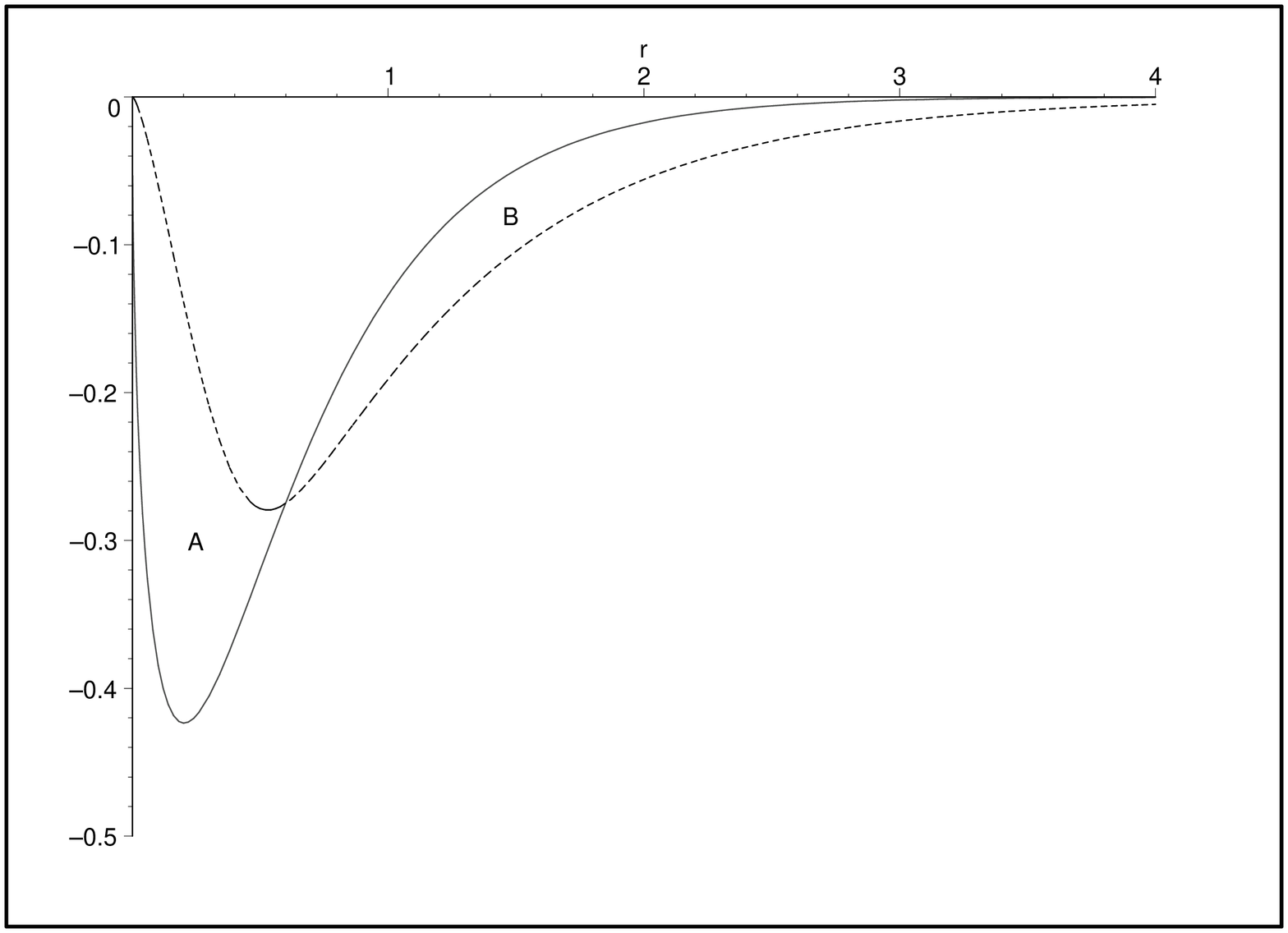}
\end{array}$
\end{center}
\caption{Left graph: potential $V_a=-\cfrac{\alpha}{(r^q + a^q)^{1/q}}$ dashed lines and $V_b=-\cfrac{\beta}{e^{r/b} - 1}$ full line. Right graph: function $V_a\psi_br$ dashed lines and $V_b\psi_br$ full line. Where the following values were applied: $\alpha=0.8$, $a=0.5$, $q=5$, $\beta=1.39$, and $b=0.8$.}
\end{figure}
Then area $A=0.105507$ and $B=0.10551$ and $\zeta(\infty)>0$ so we conclude $E_a\le E_b$. Calculating accurate numerical eigenvalues, we verify that $E_a=0.31123\le E_b=0.39008$.

\subsection{The case $V_a=V_b$ and $S_a\ne S_b$}
We assume that as before the discrete energy eigenvalue $E$ is such that $0\le E\le m$, the vector potential $V$ satisfies $1^\circ$--$\ 3^\circ$, and the scalar potential $S$ satisfies either
\begin{eqnarray*}
&7^\circ& S\ \text{is nonnegative}\ \text{i.\,e.} \  S\ge 0;\\
\vspace{7mm}
&8^\circ&  S\ \text{is nonincreasing on} \  [0, \infty), \text{so} \  S'\le 0;\\
\vspace{7mm}
&9^\circ& S\ \text{vanishes at infinity, thus}\ \lim_{r\to \infty}S=0.\phantom{-}\phantom{-}\phantom{-}\phantom{-}\phantom{-}\phantom{-}
\phantom{-}\phantom{-}\phantom{-}\phantom{-}\phantom{.}
\end{eqnarray*}
or
\begin{eqnarray*}
&10^\circ& S\ \text{is nonpositive and bounded by}\ -m,\ \text{i.\,e.} \  -m\le S\le 0;\\
\vspace{7mm}
&11^\circ&  S\ \text{is nondecreasing on} \  [0, \infty), \text{so} \  S'\ge 0;\\
\vspace{7mm}
&12^\circ& S\ \text{vanishes at infinity, thus}\ \lim_{r\to \infty}S=0.
\end{eqnarray*}

Now if two comparison vector potentials $V_a$ and $V_b$ are equal but the scalar potentals $S_a$ and $S_b$ are different, relation (\ref{rel}) can be rewritten as
\begin{equation}\label{rel_5}
(E_b-E_a)\int_0^\infty W_1\psi_a\psi_bdr=\int_0^\infty (S_b-S_a)W_2\psi_a\psi_bdr,
\end{equation}
where
\begin{equation*}
W_1=E_a+E_b-2V, \ V=V_a=V_b, \quad \text{and} \quad W_2=S_a + S_b+2m.
\end{equation*}
Then the following comparison theorem immediately follows:

\medskip

\noindent{\bf Theorem 8:} ~~{\it If $V_a=V_b$ and $S_a\le S_b$ then $E_a\le E_b$, where the vector potential $V$ satisfies $1^\circ$--$\ 3^\circ$ and the scalar potential $S$ satisfies either $7^\circ$--$\ 9^\circ$ or $10^\circ$--$\ 12^\circ$.}

\subsubsection*{An Example}
Consider the soft--core potential \cite{softcore_1, softcore_2} 
\begin{equation*}
V_a=V_b=-\cfrac{v}{\left(r^{p}+c^{p}\right)^{1/p}},
\end{equation*} 
with $v=1$, $c=2$, and $p=4$. For the scalar potentials we choose the soft--core potential $S_a$ and the sech--squared potential $S_b$:
\begin{equation*}
S_a=-\cfrac{\alpha}{\left(r^{q}+a^{q}\right)^{1/q}} \qquad \text{and} \qquad 
S_b=-\cfrac{4\beta}{\left(e^{br}+e^{-br}\right)^2}.
\end{equation*}
If $\alpha=0.8$, $a=1.6$, $q=3$, $\beta=0.5$, and $b=0.31$ the potentials are ordered $S_a\le S_b$. The condition $-m\le S\le 0$ is also satisfied for $m=1$. Then, as stated in Theorem 8, we conclude $E_a\le E_b$, which is verified by accurate numerical eigenvalues $E_a=0.50535\le E_b=0.52131$, for $d=3$.

\subsection{The case $V_a\ne V_b$ and $S_a=S_b$}
In the case $S_a=S_b$ and $V_a\ne V_b$ relation (\ref{rel}) becomes
\begin{equation}\label{spcase}
(E_b-E_a)\int_0^\infty W\psi_a\psi_bdr 
=\int_0^\infty(V_b-V_a)W\psi_a\psi_b dr
\end{equation}
where 
\begin{equation*}
W=E_a+E_b-V_a-V_b.
\end{equation*}
Then the comparison theorem follows:

\medskip

\noindent{\bf Theorem 9:} ~~{\it If $S_a=S_b$ and $V_a\le V_b$ then $E_a\le E_b$, where the vector potential $V$ satisfies $1^\circ$--$\ 3^\circ$ and the scalar potential $S$ satisfies either $7^\circ$--$\ 9^\circ$ or $10^\circ$--$\ 12^\circ$.} 

\medskip

In order to sharpen comparison theorem 9 in $d$ dimensions, we prove the following lemma, which requires that in $d=2$ dimensions if $\lim\limits_{r\to 0^+}V=V_0$, then $V_0<-m$: 

\medskip

\noindent{\bf Lemma 4:} ~~{\it The Klein--Gordon $s$--state energy eigenfunction $\psi$, which satisfies (\ref{KGVS_1}), is such that
\begin{equation*}
\left( \frac{\psi}{r^{\frac{d-1}{2}}} \right)'\le 0,
\end{equation*}
where $V$ satisfies $1^\circ$--$\ 3^\circ$ and $S$ satisfies $10^\circ$--$\ 12^\circ$.} 

\medskip

\noindent{\bf Proof:} The proof is similar to the proofs of lemmas 2 and 3; the function $F$ in that case is $F=(m+S)^2-(E-V)^2$. Assumptions $1^\circ$--$\ 3^\circ$ on $V$ and $10^\circ$--$\ 12^\circ$ on $S$ ensure that $F'\ge 0$ and there is one point $\hat{r}>0$ satisfying $F(\hat{r})=0$.

\hfill $\Box$

With the help of the above lemma and comparison relation (\ref{spcase}), we establish the following two sharpened comparison theorems:

\medskip

\noindent{\bf Theorem 10:} ~~{\it If $S$ satisfies $10^\circ$--$\ 12^\circ$, $V$  satisfies $1^\circ$--$\ 3^\circ$, has $t^{d-1}$--weighted area, and   
\begin{equation*}
\chi(r)=\int_0^r (V_b(t)-V_a(t)) t^{d-1} dt\ge 0, \quad r\in[0,\ \infty).
\end{equation*}
then $E_a\le E_b$.} 

\medskip

\noindent{\bf Theorem 11:} ~~{\it If $S$ satisfies $10^\circ$--$\ 12^\circ$, $V$  satisfies $1^\circ$--$\ 3^\circ$, has $\psi_jt^{\frac{d-1}{2}}$--weighted area, and 
\begin{equation*}
\lambda(r)=\int_0^r (V_b(t)-V_a(t))\psi_j(t) t^{\frac{d-1}{2}} dt\ge 0, 
\quad r\in[0,\ \infty),
\end{equation*}
then $E_a\le E_b$, where $j$ is either $a$ or $b$.}

\medskip

We note that for the theorems 10 and 11 corresponding corollaries for the case of one and two intersections can be stated and then generalized for the $n$ intersections.

\subsubsection*{An Example}
To demonstrate theorem 9 we take for the scalar potential $S$ the soft--core potential 
\begin{equation*}
S_a=S_b=-\cfrac{v}{(r^q+c^q)^{1/q}},
\end{equation*}
with $v=1$, $c=2$, and $q=4$. For the vector potentials we take the laser--dressed potential $V_a$ and the Woods--Saxon potential $V_b$ \cite{WS}:
\begin{equation*}
V_a=-\cfrac{\alpha}{(r^2+a^2)^{1/2}} \qquad \text{and} \qquad 
V_b=-\cfrac{\beta}{1+e^{\frac{r-R}{b}}}.
\end{equation*}
With the following values of parameters $\alpha=0.6$, $a=0.6$, $\beta=1$, $b=0.9$, and $R=0.96$ the comparison potentials are ordered, i.\,e. $V_a\le V_b$ (Figure 6, left graph) and Theorem~9 predicts the energy ordering  $E_a\le E_b$. Meanwhile we find numerically that $E_a=0.54094\le E_b=0.60543$, for $d=3$.
\begin{figure}[ht]
\begin{center}$
\begin{array}{cc}
\includegraphics[width=2.5in, angle=-90]{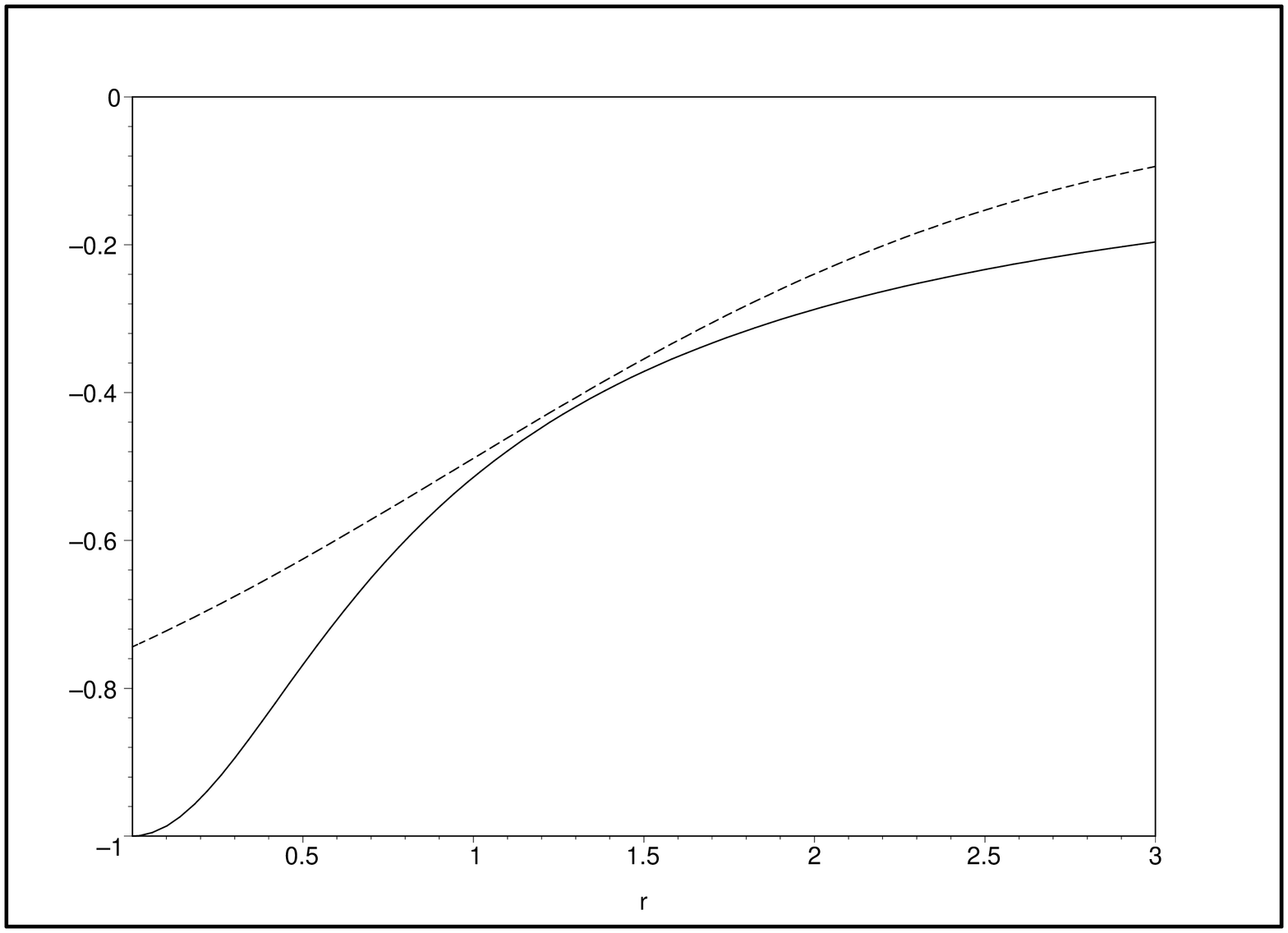} & 
\includegraphics[width=2.5in, angle=-90]{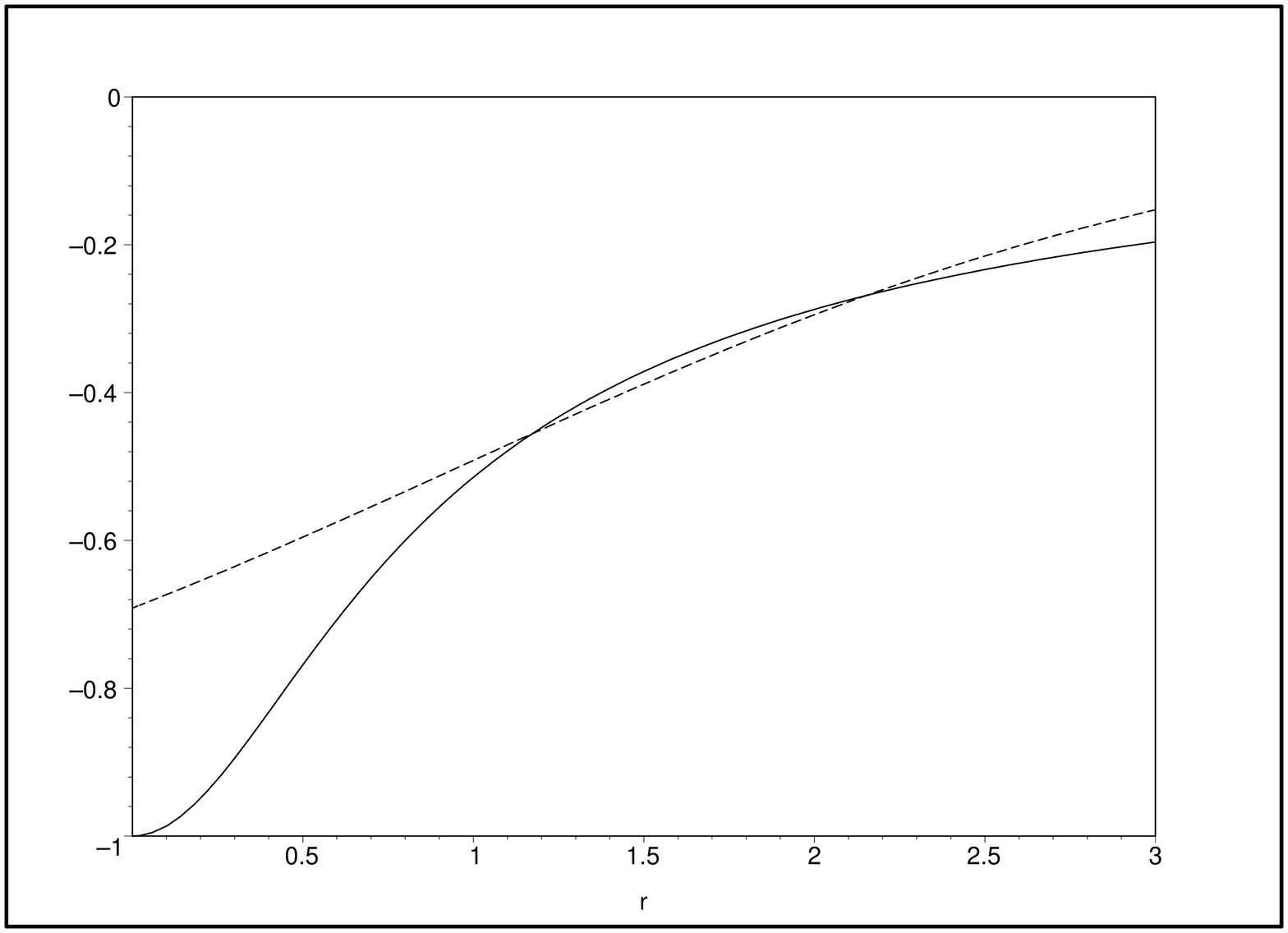}
\end{array}$
\end{center}
\caption{Left graph: potential $V_a=-\cfrac{\alpha}{(r^2+a^2)^{1/2}}$ full line and $V_b=-\cfrac{\beta}{1+e^{\frac{r-R}{b}}}$ dashed lines, where $\alpha=0.6$, $a=0.6$, $\beta=1$, $b=0.9$, and $R=0.96$. Right graph: potential $V_a=-\cfrac{\alpha}{(r^2+a^2)^{1/2}}$ full line and $V_b=-\cfrac{\beta}{1+e^{\frac{r-R}{b}}}$ dashed lines, where $\alpha=0.6$, $a=0.6$, $\beta=1$, $b=1.19$, and $R=0.96$.}
\end{figure}

Now let us sharpen the previous result. Using theorem 10, we write the corollary for the case of two intersections: {\it If the graphs of the comparison potentials cross over twice, at $r=r_1$ and $r=r_2$, $r_1<r_2$, $V_a\le V_b$ for $r\in [0,\ r_1]$, and
\begin{equation*}
\chi(r_2)=\int_0^{r_2} (V_b-V_a)r^{d-1} dr\ge 0, \quad \text{then}\quad E_a\le E_b.  
\end{equation*}} 
If we change the value of $b$ from $0.9$ to $1.19$ in the comparison potential $V_b$, then the potentials intersect twice, with $V_a\le V_b$ before the first intersection point: see Figure 6, right graph.  The value $\chi(r_2)\ge 0$, thus we should have $E_a\le E_b$.  By numerical calculations we find $E_a=0.54094\le E_b=0.56950$. As we can see the difference between the comparison eigenvalues is smaller in the sharp case.
\section{Conclusion}
In order to exhibit the theorem differences we have first re-derived the basic comparison theorems \cite{HallComp, Cui}. By using established properties of the nodeless radial wave functions, we are able to sharpen these basic theorems so that the graphs of the the comparison potentials are allowed to intersect each other whilst the corresponding eigenvalues remain ordered. In fact, we replace the inequality $V_a\le V_b$, by the weaker assumption $U_a\le U_b$, where, for $d=1$, $U_i(x)=\int_0^x V_i(t)dt$, $i=a$ or $b$, and for $d>1$, $U_i(r)=\int_0^r V_i(t) t^{d-1}dt$, $i=a$ or $b$. Thus the new condition $U_a\le U_b$ leads to the spectral ordering $E_a\le E_b$ as well. Even weaker sufficient conditions for spectral ordering are possible if one of the comparison problems has a known wave function for it may be used to further sharpen the energy bounds.  The reason for this is that the the difference $V_b-V_a$ is multiplied by the wave function, which decreases monotonically to zero at infinity, thus allowing even wider potential intersections to imply spectral ordering.. We have derived a number of corollaries which make the application of the theorems very straightforward.

\medskip
In $d > 1$ dimensions, the sharp comparison theorems are first established for the ground-state eigenvalue $E_{00}^d$.  
The  equivalence theorem 3, to the effect that $E_{0\ell}^d=E_{00}^{d+2l}$, then allows us to apply the 
 the results for  $E_{00}^d$ to $E_{0\ell}^{d'}$, where $d = d'+2\ell$.

\medskip
In $d>1$ dimensions we have also proved a variety of  sharp comparison theorems in the presence of a scalar potential $S$. In particular, sharp comparison theorems in the spin--symmetric $S=V$, and pseudo--spin--symmetric $S=-V$ cases were derived. Similarly, sharp comparison theorems in the special cases $V_a=V_b$ while $S_a\ne S_b$, and $V_a\ne V_b$, while $S_a=S_b$, are discussed.  It is clear that very similar theorems may be established {\it mutatis mutandis} in $d=1$ dimension.

\section{Acknowledgments}
One of us (RLH) gratefully acknowledges partial financial support
of this research under Grant No.\ GP3438 from the Natural Sciences
and Engineering Research Council of Canada.\medskip

\section*{References}


\begin{thebibliography}{99}
\bibitem{HallComp} R. L. Hall and M. D. Aliyu, Phys. Rev. A {\bf 78}, 052115 (2008).
\bibitem{Cui} C.--B. Luo, C.--Y. Long, Z.--W. Long, and S.--J. Qin, Physics Letters A {\bf 376}, 703 (2012).
\bibitem{Reed} M. Reed and B. Simon, {\it Methods of Modern Mathematical Physics IV: Analysis of Operators}, (Academic, New York, 1978). 
\bibitem{Thirring} W. Thirring, {\it A Course in Mathematical Physics 3: Quantum Mechanics of Atoms and Molecules}, (Springer, New York, 1981). 
\bibitem{Fr} J. Franklin and L. Intemann, Phys. Rev. Lett. {\bf 54}, 2068 (1985).
\bibitem{Gold} S. P. Goldman, Phys. Rev. A {\bf 31}, 3541 (1985).
\bibitem{Gr} I. P. Grant and H. M. Quiney, Phys. Rev. A {\bf 62}, 022508 (2000).
\bibitem{Hall1} R. L. Hall, Phys. Rev. Lett. {\bf 83}, 468 (1999).
\bibitem{Hall2} R. L. Hall, Phys. Rev. A {\bf 81}, 052101 (2010).
\bibitem{monoton1} R. L. Hall, Phys. Rev. Lett. {\bf 101}, 090401 (2008).
\bibitem{monoton2} R. L. Hall and M. D. Aliyu, Phys. Rev. A {\bf 78}, 052115 (2008).
\bibitem{monoton3} R. L. Hall and \"{O}. Ye\c{s}ilta\c{s}, J. Phys. A: Math. Theor. {\bf 43}, 195303 (2010).
\bibitem{Greiner} W. Greiner, {\it Relativistic Quantum Mechanics: Wave Equations}, third ed., (Springer, Berlin, 2000). The square well is discussed on pp. 56--59.
\bibitem{Barton} G. Barton, J. Phys. A: Math. Gen. {\bf 40}, 1011 (2007).
\bibitem{HallCO} R. L. Hall, Phys. Lett. A {\bf 372}, 12 (2007).
\bibitem{Hall3} R. L. Hall, J. Phys. A {\bf 25}, 4459 (1992).
\bibitem{ddimSch} R. L. Hall and Q. D. Katatbeh, J. Phys. A {\bf 35}, 8727 (2002).
\bibitem{Hall4} R. L. Hall and N. Saad, Phys. Lett. A {\bf 237}, 107 (1998).
\bibitem{KGd1_1} P. Strange, {\it Relativistic Quantum Mechanics}, (Cambridge University Press, 1998).
\bibitem{Ref} R. L. Hall, J. Phys. A: Math . Gen. {\bf 25}, 4459 (1992).
\bibitem{cutoff_2} Mehta C. H. and Patil S. H., Phys. Rev. A {\bf 17}, 43 (1978).
\bibitem{cutoff_1} R. L. Hall and Q. D. Katatbeh, Phys. Lett. A. {\bf 294}, 163 (2002).
\bibitem{sechsquared_4} P. S. Epstein, Proc. Natl. Acad. Sci. U.S.A. {\bf 16}, 627 (1930).
\bibitem{sechsquared_2} C. Eckart, Phys. Rev. {\bf 35}, 1303 (1930).
\bibitem{sechsquared_1} G. P\"{o}schl and E. Teller, Z. Phys. {\bf 83}, 143 (1933).
\bibitem{sechsquared_3} B. Y. Tong, Solid State Commun. {\bf 104} (11), 679 (1997).
\bibitem{Alhaidari} A. D. Alhaidari, H. Bahlouli, A. Al--Hasan, Phys. Lett. A {\bf 349}, 87 (2006).
\bibitem{eval} R. Friedberg, T. D. Lee, and W. Q. Zhao, Ann. Phys. {\bf 321}, 1981 (2006).
\bibitem{Nieto} M. M. Nieto, Am. J. Phys. {\bf 47}, 1067 (1979).
\bibitem{Col1} Z.--Q. Ma, S.--H. Dong, Z.--Y. Gu, J. Yu, and M. Lozada--Cassou, Int. J. Mod. Phys. E {\bf 13}, 597 (2004).
\bibitem{Col2} N. Saad, R. L. Hall, and H. Ciftci, Cent. Eur. J. Phys. {\bf 6}, 717 (2008). 
\bibitem{softcore_1} R. L. Hall, N. Saad, K. D. Sen, and H. Ciftci, Phys. Rev. A. {\bf 80}, 032507 (2009).
\bibitem{softcore_2} R. L. Hall, N. Saad, and K. D. Sen, J. Math. Phys. {\bf 51}, 022107 (2010).
\bibitem{Yuk} H. Yukawa, Proc. Phys. Math. Soc. Japan {\bf 17}, 48 (1935).
\bibitem{Hult_1} L. Hulth\'{e}n, Ark. Mat. Astron. Fys. {\bf 28}A, 5 (1942).
\bibitem{Hult_2} L. Hulth\'{e}n, M. Sugawara, and S. Flugge (ed.), {\it Handbuch
der Physik}, (Springer, Berlin, 1957).
\bibitem{Domingues} F. Dominguez--Adame, Phys. Lett. A {\bf 136}, 175 (1989).
\bibitem{WS} R. D. Woods and D. S. Saxon, Phys. Rev. {\bf 95}, 577 (1954).
\end{thebibliography}
\end{document}